\documentclass[conference]{IEEEtran}
\IEEEoverridecommandlockouts

\usepackage[utf8]{inputenc} % According to the encoding used, you may replace "utf8" with "latin5"
\usepackage[T1]{fontenc}

\usepackage{footmisc}

\usepackage[pdftex]{graphicx}

\usepackage[tight,footnotesize]{subfigure}

\usepackage{cite}

\usepackage{soul}

\usepackage[cmex10]{amsmath}
\usepackage{amssymb}
\usepackage{mathrsfs}
\usepackage{siunitx}

\usepackage{url}
\usepackage{lipsum}
\usepackage{multirow}
\usepackage{array}
\usepackage{gensymb}

\newcolumntype{L}[1]{>{\raggedright\let\newline\\\arraybackslash\hspace{0pt}}m{#1}}
\newcolumntype{C}[1]{>{\centering\let\newline\\\arraybackslash\hspace{0pt}}m{#1}}
\newcolumntype{R}[1]{>{\raggedleft\let\newline\\\arraybackslash\hspace{0pt}}m{#1}}

\usepackage{color}
\usepackage{todonotes}
\definecolor{morange}{rgb}{0.8,0.2,0}
\definecolor{mblue}{rgb}{0,0.3,1.0}
\definecolor{mbluee}{rgb}{0.4,0.1,0.9}
\definecolor{mpink}{rgb}{0.8,0.3,0.4}
\definecolor{mgreen}{rgb}{0.2,0.4,0}

\begin{document}

\title{Uncertainty Quantification in Molecular Signals using Polynomial Chaos Expansion
\thanks{The work is partially funded by the US AFOSR grant FA9550-17-1-0056, EPSRC grant EP/R041725/1, and the Alan Turing Institute under the EPSRC grant EP/N510129/1 and the Data-Centric Engineering Program funded by Lloyd's Register Foundation. $^*$Corresponding Author: weisi.guo@warwick.ac.uk}
}

\author{
\IEEEauthorblockN{Mahmoud Abbaszadeh$^{1}$, Giannis Moutsinas$^{1}$, Peter J. Thomas$^{1}$, Weisi Guo$^{1,2*}$\\} %\vspace{0.1cm}
\IEEEauthorblockA{
$^{1}$School of Engineering, University of Warwick, Coventry, United Kingdom\\
$^{2}$Alan Turing Institute, London, United Kingdom \\
}
}

\maketitle

\begin{abstract} % Weisi
Molecular signals are abundant in engineering and biological contexts, and undergo stochastic propagation in fluid dynamic channels. The received signal is sensitive to a variety of input and channel parameter variations. Currently we do not understand how uncertainty or noise in a variety of parameters affect the received signal concentration, and nor do we have an analytical framework to tackle this challenge. In this paper, we utilize Polynomial Chaos Expansion (PCE) to show to uncertainty in parameters propagates to uncertainty in the received signal. In demonstrating its applicability, we consider a Turbulent Diffusion Molecular Communication (TDMC) channel and highlight which parameters affect the received signals. This can pave the way for future information theoretic insights, as well as guide experimental design.
\end{abstract}


\section{Introduction} % Weisi
Molecular signals are abundant in biological signaling \cite{Wyatt14, unluturk2017end, farsad2016comprehensive}, industrial engineering (e.g. chemical catalysis \cite{Icardi14}), nano-engineering \cite{Akyildiz15}, and ecosystems (e.g. pollution signals in rivers \cite{Sonnenwald14}). In many cases, the signals represent explicit information (e.g. pheromones \cite{Wyatt14}), and in other cases the chemical plume patterns are a proxy signal for an opaque process (e.g. chemical mixing-reaction rate \cite{Guida10}). Molecular signals exist in different scales, from nano-scale diffusion dynamics in cells to macro-scale flow in oceans. In all cases, identical molecular signal plumes never give rise to the same statistical channel response due to external disturbances and other uncertainties. Yet, simulating all possible permutations and considering all fluid dynamic forces is expensive and there is a need to quantify uncertainty more directly.

\subsection{Uncertainty in Molecular Signal Propagation} % Weisi
Uncertainty can arise from noise in the input and ambient parameters. This makes deterministic models unreliable in estimating the variational behaviour of complex systems. Many engineering systems consist of complex differential equation models, which give deterministic outputs. In the case of mass diffusion based molecular signaling (and molecular communication), the classic Fick's Law yields a deterministic inverse-Gaussian form \cite{Guo16WC}. When considering more complex forces (e.g. turbulence, sheer stress) \cite{unluturk2017end, Rose18, Abbaszadeh18VR}, the Reynolds-Averaged-Navier-Stokes (RANS) equations still yield deterministic solutions. This means that externally triggered variations in input parameters (e.g. velocity profile of molecular signal) and the channel parameters (e.g. dynamic viscosity, diffusivity) cannot be accounted for. Monte-Carlo simulations are required to simulate variational behaviour, causing time intensive computation and lacking in direct insight about sensitivity to different parameter combinations.

\subsection{Review of Similar Work}
Monte-Carlo simulation, whilst time consuming, can offer computation convergence guarantees in the face of multiple uncertainties \cite{mathelin2005stochastic}. Often, as is the case for fluid dynamics, there are divergent solutions to the model. In weather forecasting, extreme weather represents one of the many possible outcomes and uncertainty propagation is essential. In \cite{farazmand2017variational}, the authors use sparse initial weather data to inform the likelihood of divergent solutions forming. In probabilistic programming and numerics \cite{Girolami15}, uncertainty is cascaded through to yield posterior estimates of the solution, which is computationally expensive for simulations. Nonetheless, the aforementioned scenarios are computationally expensive and data demanding. As such, analytical methods for uncertainty propagation in fluid dynamics is useful.

Polynomial chaos expansion (PCE) is a method to determine the propagation of uncertainty in dynamical systems, when there is probabilistic uncertainty in the system parameters. PCE has been widely used in fluids dynamics since faithful molecular dynamic or finite-element computation is usually time-consuming. Hosder et al. \cite{hosder2006non} employed the non-intrusive PCE to add uncertainty in the inputs of aerodynamic forces where the uncertainty was in the geometry of the channel, and laminar boundary layer flow with the uncertain parameter of dynamic viscosity. For evaluation, they compared their results with Monte Carlo simulation and a good agreement existed. The main benefit is that PCE can achieve the same results with a 7-order computational saving \cite{hosder2006non}. 

In molecular signaling, when the propagation channel is described by a stable mass diffusion equation, the uncertainty arises from Brownian motion \cite{srinivas2012molecular}. However, when there are fluctuations in the diffusion channel from temperature and diffusivity variations, then uncertainty in the channel propagates to the receiver signal concentration \cite{qiu2017molecular}. As evidenced by the literature, there is a lack of uncertainty research in molecular communications under complex fluid dynamic forces. We know this will inform our knowledge of new noise sources (e.g. transposition noise \cite{Haselmayr17}) and the information capacity. Therefore, PCE represents a good avenue of research. It is worth noting that even in the whole area of wireless communication, there is a lack of research in uncertainty propagation \cite{acikgoz2017statistical, boeykens2014efficient}. 

\subsection{Contribution and Organization of Paper} % Weisi
In Section II, polynomial chaos expansion from a general point of view has been introduced to quantify uncertainty in dynamical systems. In Section III, we introduce the molecular signal channel, including the RANS equation and integration with PCE. In Section IV, we present the results and discuss applications of PCE.

%%%%%%%%%%%%%%%%%%%%%%%%%%%%%%%%%%%%%%%%%%%%%%%%%%
\section{Polynomial Chaos Expansion}
PCE is a method which facilitates the spectral representation of the uncertainty in physics-based and engineering problems. In this surrogate method, the output can be represented as a series of the input random parameters so the uncertainty in the input parameters would be reflected in the outputs \cite{owen2017comparison}. Each input is considered as a random variable with a specific probability density function (PDF) and the goal of this method is to find a function which relates the random inputs to the random output as a series.      
%%%%%%%%%%%%%%%%%%%%%%%%%%%%%%%%%%%%%%%%%%%%%%%%%%
\subsection{Univariate Polynomial Chaos}
Let $\Xi$ be a random variable with known PDF $w$, and $X=\phi(\Xi)$, where $\phi$ is a function that is square integrable on $\chi$ ($\chi\subset\mathbb{R}$) with $w$ as a weight function (call this space $L_w^2$). Our goal is to approximate $X$ by a polynomial series of $\Xi$. For this purpose, we need a family of polynomials $P_n$ such that $P_0$ is not $0$, and for $n\ge0$ ($n\in\mathbb{N}$), the polynomial $P_n$ has the order of $n$ and are orthogonal with respect to $w$. Thus,
\begin{equation}\begin{split}
\label{e1}
 \langle P_n, P_m \rangle_w = \int_{\chi} P_m(x) P_n(x) w(x) dx=\gamma_m\delta_{mn},
\end{split}\end{equation} where $\delta_{mn}$ is the Kronecker delta and would be 1 if $m=n$ and 0 if $m\neq n$, and $\gamma_m$ is the normalization constant which is obtained by:
\begin{equation}\begin{split}
\label{e2}
 \gamma_m\equiv \int_{\chi} P^2_m(x) w(x) dx.
\end{split}\end{equation} We also assume that $P_0$ is normalized so that $\langle P_0,P_0 \rangle_w = 1$. Depending on the distribution of the $\Xi$, different set of orthogonal polynomials should be employed to satisfy \eqref{e1}. 

%%%%%%%%%%%%%%%%%%%%%%%%%%%%%%%%%%%%%%%%%%%%%%%%%%%%%%%%%
\subsubsection{Legendre Polynomials}
If $\Xi$ has a uniform distribution ($\Xi \sim$ Unif[-1,1]), Legendre polynomials should be used \cite{xiu2003modeling}. By knowing the first two Legendre polynomials, which are $P_0(x)=1$ and $P_1(x)=x$, we can generate the rest by using the following recursive relation \cite{chihara2011introduction}:
\begin{equation}\begin{split}
\label{e3}
 (m+1)P_{m+1}(x)=(2m+1)xP_m(x)-mP_{m-1}(x),
\end{split}\end{equation} where $m$ is the order of the polynomial. The generated polynomials are orthogonal if we consider $w(x)=\frac{1}{2}$ which is the PDF of $\Xi \sim$ Unif[-1,1]. In this case, the normalization constant for $m\in \mathbb{N}$ is $1/(2m+1)$.

%%%%%%%%%%%%%%%%%%%%%%%%%%%%%%%%%%%%%%%%%%%%%%%%%%%%%%%%% 
\subsubsection{Hermite polynomials}
When $\Xi$ has a normal distribution ($\Xi \sim N(0,1)$), then the Hermite polynomials should be employed to build the PCE \cite{xiu2003modeling}. By considering $P_0(x)=1$, the recursive relation for Hermite polynomials is \cite{chihara2011introduction}:
\begin{equation}\begin{split}
\label{e4}
 P_{m+1}(x)=xP_m(x)-\frac{d}{dx}P_{m}(x).
\end{split}\end{equation} The generated polynomials are orthogonal if we consider $w(x)=\frac{\exp(-x^2/2)}{\sqrt{2\pi}}$ which is the PDF of $\Xi \sim N(0,1)$. In this case, the normalization constant for $m\in \mathbb{N}$ is $m!$.

There are also other PDFs and their corresponding polynomials in the literature that can be considered based on the type of the input variables \cite{xiu2003modeling}. In this study, we only used uniform and normal distributions as PDF of the uncertain input parameters in molecular signaling. 

Now, the series can be built by considering that the polynomials $P_n$ can be used as a basis for $L_w^2$.
\begin{equation}\begin{split}
\label{e5}
  \phi(\Xi)=\sum_{n\ge0}a_n P_n(\Xi).
\end{split}\end{equation}
where $a_n$ are deterministic unknown coefficients. Because $P_n$ is an orthogonal basis, the coefficients by projecting on each basis vector.
\begin{equation}\begin{split}
\label{e6}
  a_n = \frac{\langle \phi, P_n\rangle_w}{\langle P_n, P_n \rangle_w}.
\end{split}\end{equation}

After calculating the $a_n$ coefficients, the statistics of the output $X$ can be determined spatially. Due to the orthogonality of the applied polynomials, the expectation of the $X$ is estimated by:
\begin{equation}\begin{split}
\label{e7}
  E[X] \approx \alpha_0,
\end{split}\end{equation} 
which is the coefficient of the zeroth order polynomial. Also, the variance can be estimated in the same way.
\begin{equation}\begin{split}
\label{e8}
  \text{Var}[X] \approx \sum_{n\ge1}\alpha_n^2 \langle P_n,P_n \rangle_w. 
\end{split}\end{equation}

Since we cannot simplify a product of three or more polynomials, the equations for further moments contain all possible terms of the power of the sum.

%%%%%%%%%%%%%%%%%%%%%%%%%%%%%%%%%%%%%%%%%%%%%%%%%%%%%%%%%%%%%%
\subsection{Parametric Univariate Polynomial Chaos}

The case where $X$ depends also on an independent parameter, $t$, can be treated as a straightforward generalization of the previous. In particular if $X=\phi(t,\Xi)$ with $\phi(t,\cdot) \in L_w^2$ for all $t$, then we can decompose $X$ as 
\begin{equation}\begin{split}
\label{e9}
 \phi(t,\Xi)=\sum_{n\ge0}a_n(t) P_n(\Xi). 
\end{split}\end{equation}

This becomes particularly useful in the case of differential equations that depend on a stochastic parameter. For example, in the case of a linear differential equation $\dot{X}=L(X)+\psi(\Xi)$, where $L$ is linear and $\psi(\Xi) = \sum_{n\ge 0}b_n P_n(\Xi) $, we can substitute the sum and use the linearity of the equation to get
\begin{equation}\begin{split}
\label{e10}
 \sum_{n\ge0}\dot{a}_n(t) P_n(\Xi) = \sum_{n\ge0} L(a_n(t)) P_n(\Xi) + \sum_{n\ge 0}b_n P_n(\Xi).
\end{split}\end{equation}
and by projecting on each $P_n$ we get the equations $\dot{a}_n = L(a_n) + b_n$, whose solutions are the coefficients of the expansion. Notice that all the equations now are deterministic.

In the case of non-linear differential equation, one has products of polynomial chaos expansions. This means that one gets an infinite system of deterministic differentiable equations. This method of computing the coefficients is called \textbf{Intrusive Method} because it requires to change drastically the solver \cite{xiu2003modeling}. 

%%%%%%%%%%%%%%%%%%%%%%%%%%%%%%%%%%%%%%%%%%%%%%%%%%%%%%%%%%%%%%%%
\subsection{Multivariate Polynomial Chaos}
Let $X=\phi(\Xi_1,\Xi_2)$ where $\Xi_1$ and $\Xi_2$ are random variables with PDFs $w_1$ and $w_2$, respectively and $\phi(\cdot,\Xi_2)\in L_{w_1}^2$, $\phi(\Xi_1,\cdot)\in L_{w_2}^2$ for all $\Xi_1$ and $\Xi_2$. Let $P_{1,m}$ and $P_{2,m}$ be polynomial families that form an orthogonal basis in $L_{w_1}^2$ and $L_{w_2}^2$, respectively and $\langle P_{1,0},P_{1,0}\rangle_{w_1}=\langle P_{2,0},P_{2,0}\rangle_{w_2}=1$.
Then we can decompose $X$ as the deterministic part (coefficients) and stochastic terms. 
\begin{equation}\begin{split}
\label{e11}
  \phi(\Xi_1,\Xi_2) = \sum_{m\ge 0}\sum_{n\ge 0} a_{m,n}\, P_{1,m}(\Xi_1) P_{2,n}(\Xi_2).
\end{split}\end{equation}

One can project on the basis to get the coefficients, i.e.
\begin{equation}\begin{split}
\label{e12}
 a_{m,n} = \frac{\langle \langle \phi ,P_{1,m}\rangle_{w_1} ,P_{2,n}\rangle_{w_2}}{\langle P_{1,m},P_{1,m}\rangle_{w_1}\langle P_{2,n},P_{2,n}\rangle_{w_2}}.
\end{split}\end{equation}

Similarly to the univariate case the relations for the first two moments are relatively simple. For the expectation one has $E[X]\approx a_{0,0}$ and for the variance we have:
\begin{equation}\begin{split}
\label{e13}
Var[X] \approx \sum_{m\ge1}\sum_{n\ge1} a_{m,n}^2\, \langle P_{1,m},P_{1,m}\rangle_{w_1}\langle P_{2,n},P_{2,n}\rangle_{w_2}.
\end{split}\end{equation}

This expansion can be generalized to more than $2$ random variables in a straightforward way. The PDF of each random variable defines as inner product in the space $L_{w_i}^2$ and we have to choose an orthogonal basis in this space. Then one just expands $X$ with respect to every family of basis functions. Similarly we can generalize this scheme to treat parametric multivariate cases. One just needs to notice that the coefficients of the expansion depend only on deterministic parameters.
%%%%%%%%%%%%%%%%%%%%%%%%%%%%%%%
\begin{table}
\small
\centering
\caption{Simulation Parameters}
\label{parameters}
\begin{tabular}{ll}
\hline
Variable                             & Value \\ \hline
Maximum Injection Velocity, $u_{0}$    & \SI{4}{\meter/\second}  at $t=0$\\
Dynamic Viscosity of water, $\mu$  & \SI{8.9e-4} Pa.s\\
Density of water, $\rho$            & \SI{1000}{\kilo\gram/\metre^{3}}\\
Transmit Concentration, $c_0$       & \SI{4}{\mol/\meter^3} \\
Pulse Width, $T_0$                  & \SI{0.7}{\second}  \\
Radius of the injector ($r_{in}$)   & \SI{12.5}{\centi\meter} \\
Distance Between TX and RX, $d_{Tx,Rx}$   & $ 40\times r_{in}$\\
Simulation Dimensions                  & $50\times10\times10$ $m^3$ \\ \hline
\end{tabular}
\end{table}
%%%%%%%%%%%%%%%%%%%%%%%%%%%%%%%%%%%
%%%%%%%%%%%%%%%%%%%%%%%%%%
\begin{figure}[t]
	\centering
	\includegraphics[width=0.9\linewidth]{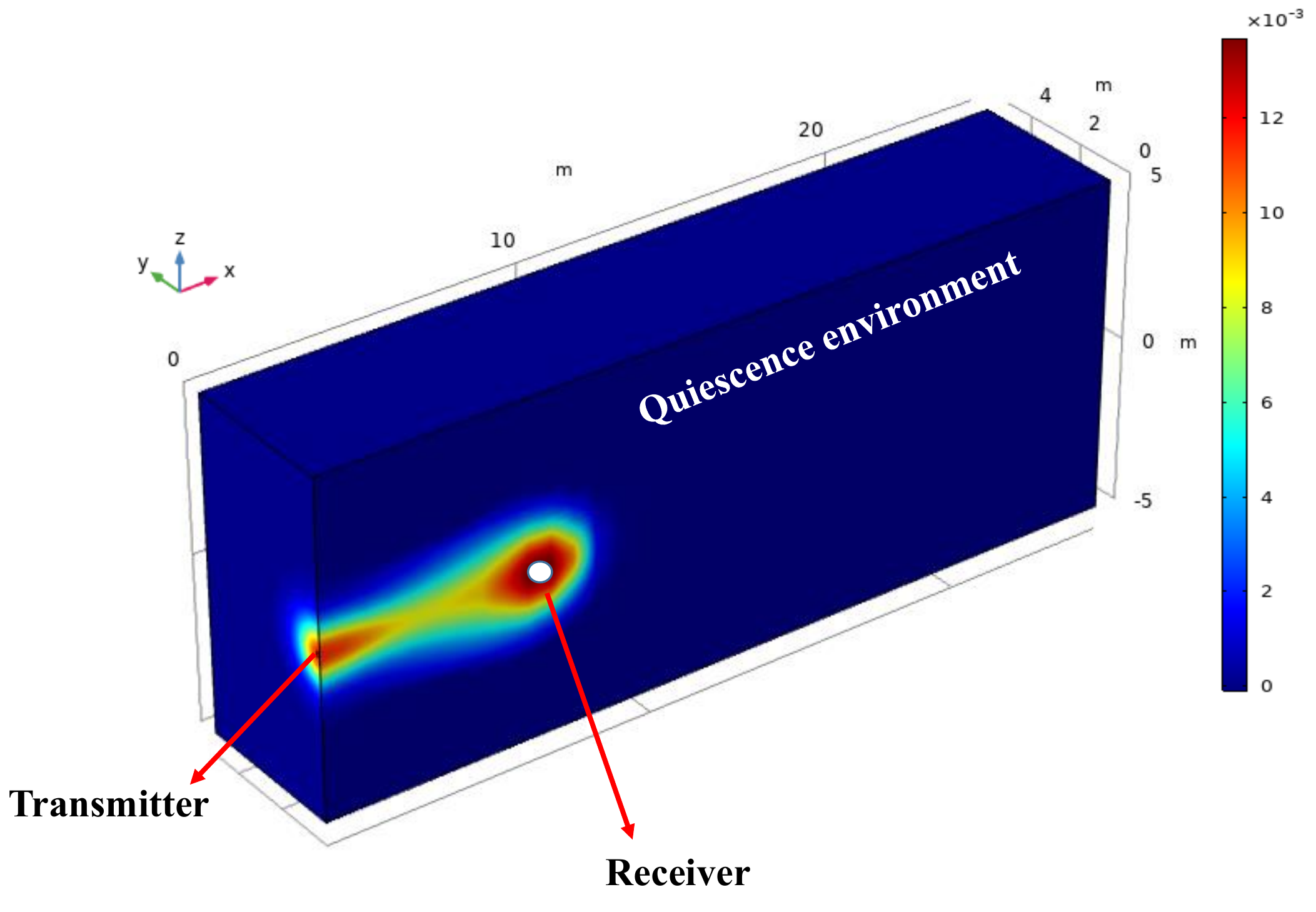}
	\caption{Schematic of the cross-section cut of the system model illustrating the quiescence environment,the transmitter, the receiver, and the emitted molecules.}
	\label{1}
\end{figure}
%%%%%%%%%%%%%%%%%%%%%%%%%%

%%%%%%%%%%%%%%%%%%%%%%%%%%%%%%%%%%%%%%%%%%%%%%%%%%%%%%%%%%%%%%%%
\subsection{Approximation using Polynomial Chaos}

In order to do any computation with a PCE series, we need to truncate it since it is not feasible to expand the PCE series to infinity. For this purpose, we notice that if the series converges, then the size of the coefficients go to $0$ if we take the limit of any index to infinity. This means that for every convergent such series we can ignore terms of order higher than some $N$. However, for a given problem it is not trivial to find which exactly this $N$ is. Usually this is done by trial and error, where we calculate more terms until the size of the new terms is smaller than the precision required. 

We start by truncating the series to an arbitrary order $N$,
\begin{equation}
    \phi_n(\Xi) = \sum_{n=0}^N a_n P_n(\Xi)
    \label{e13.5}
\end{equation}
and assume that this is enough for the considered precision. By using a truncated series, the intrusive method produces a finite system of differential equations which we can solve to get the coefficients.

There is also a \textbf{Non-intrusive Method} which which was first introduced by \cite{walters2003towards}. In this case, we observe that \eqref{e13.5} is linear with respect to $a_n$'s, so we treat the simulator as a \textbf{black box}. We generate $M\ge N$ instances of the random variable $\Xi$, $\{\xi_1,\dots,\xi_M\}$ and we calculate the deterministic outputs by simulator, $\{\phi(\xi_1),\dots,\phi(\xi_M)\}$. Then for every $\xi_i$ we have the equation:
\begin{equation}\begin{split}
\label{e14}
\phi(\xi_i) = \sum_{n=0}^N a_n P_n(\xi_i).
\end{split}\end{equation}
Notice that $\phi(\xi_i)$ and $P_n(\xi_i)$ are just numbers and now we can compute the coefficients $a_n$ by solving a linear regression. After that we compute $\sup_\Xi |a_N P_N(\Xi)|$ and if it is smaller than the precision, we stop, otherwise we increase $N$ and repeat the process. 
%%%%%%%%%%%%%%%%%%%%%
\begin{figure*}[!t]
	\centering
	\includegraphics[width=0.9\linewidth]{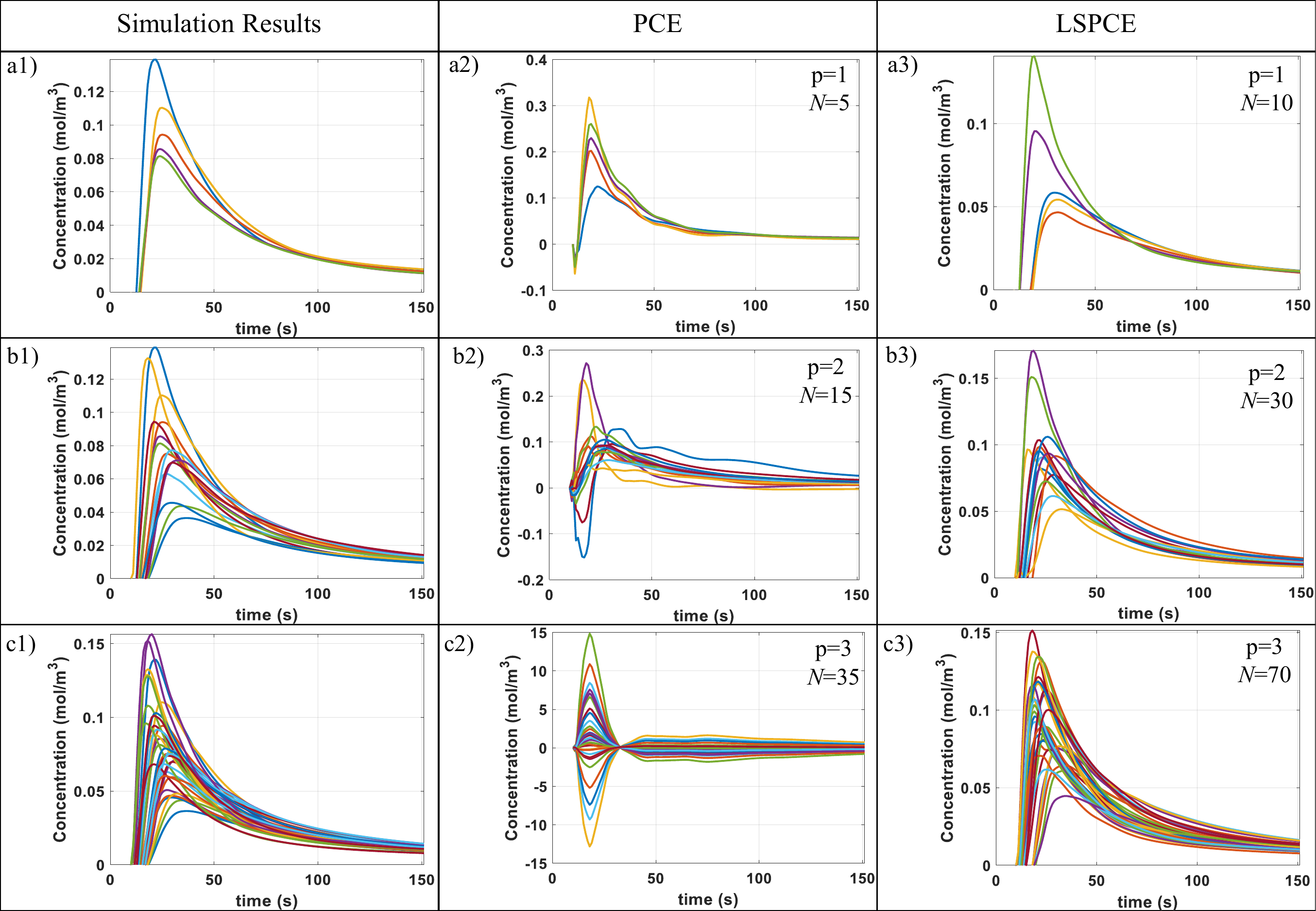}
	\caption{Concentration profiles obtained from simulation (first column), PCE (second column), and LSPCE (third column)}
	\label{2}
\end{figure*}
%%%%%%%%%%%%%%%%%%%%%
%%%%%%%%%%%%%%%%%%%%%%%%%%%%%%%%%%%%%%%%
\section{System Model and Method}
PCE is a surrogate approximation method for Monte Carlo simulation which has been widely used in communication area especially in molecular communication. In this study, to mimic reality, we employed PCE in a turbulent diffusion molecular communication channel to add uncertainty in a set of parameters such as inlet velocity, initial concentration, dynamic viscosity of the ejected fluid, and the turbulent Schmidt number (or $Sc$ which describes the ratio between the rates of turbulent transport of momentum and the turbulent transport of mass).

%%%%%%%%%%%%%%%%%%%%%%%%%%%%%%%%%%%%%%
\subsection{Channel Configuration}
The system model in this study is a 3D $50\times 10 \times 10$ $m^3$ water channel (see Fig.\ref{1}). The water molecules are ejected into a quiescence aqueous environment with \textbf{V}$=u_0i+v_0j+w_0k$ velocity where $u_0$ is taken as 4 $m/s$ and the other two components assumed to be zero $m/s$ if we have ideal injection system. The concentration of the ejected water is measured at the receiver site which is located at $40\times r_{in}$ where $r_{in}$ is the radius of the injector and the initial concentration of the water molecules is 4 $mol/m^3$. The sidewalls and the outlet has been considered far enough from the transmitter so that we can neglect their effects on the fluid flow. The properties of the water and the other system parameters are given in Table \ref{parameters}. 
%%%%%%%%%%%%%%%%%%%%%%%%%%%
\begin{figure*}[t]
	\centering
	\includegraphics[width=0.9\linewidth]{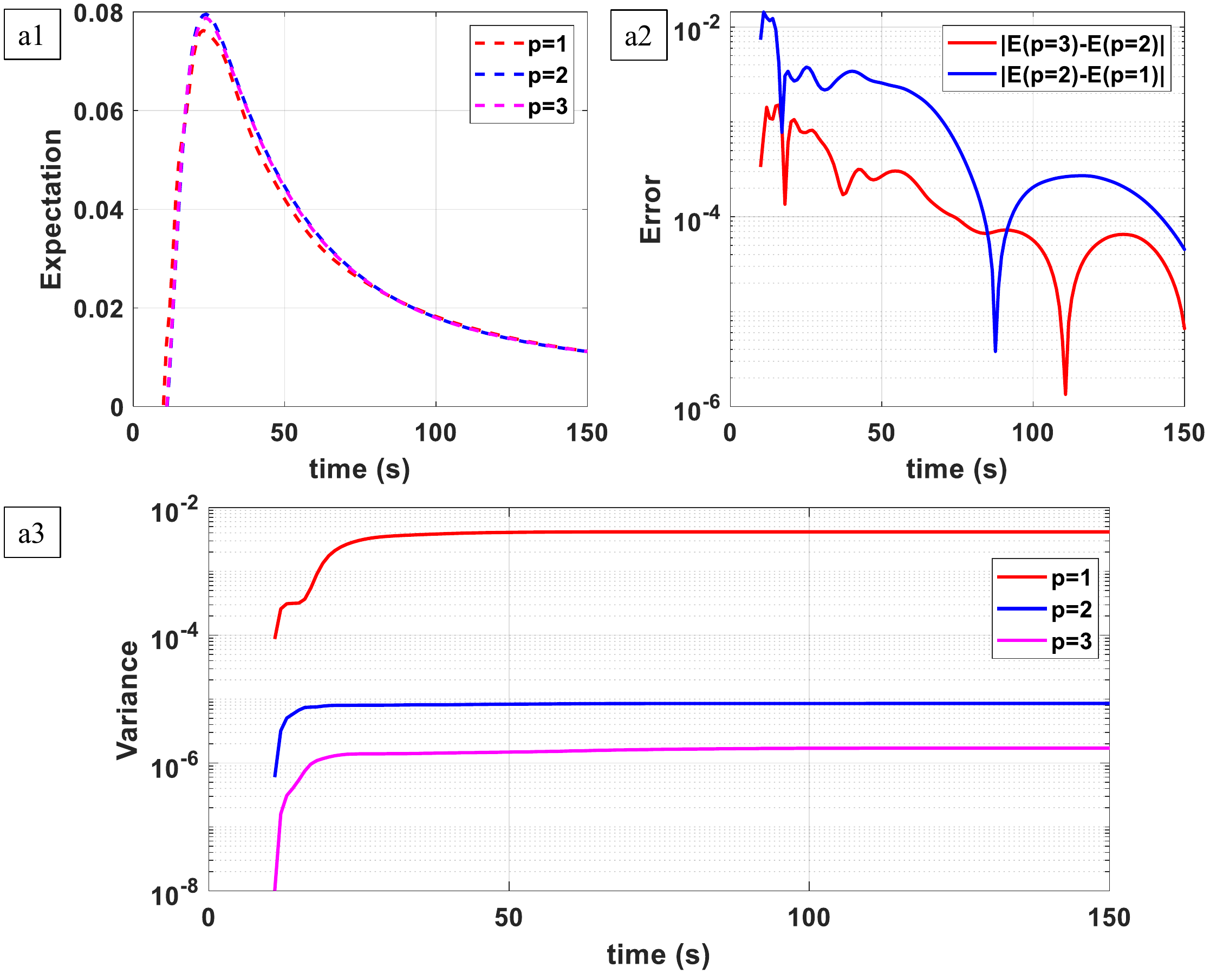}
	\caption{(a1) Expectation of LSPCE for different polynomial orders and empirical mean of concentration. (a2) Difference between the expectations by increasing the order of the applied polynomials. (a3) Variances of LSPCE for different polynomial orders.}
	\label{3}
\end{figure*}

%%%%%%%%%%%%%%%%%%%%%%%%%%%
%%%%%%%%%%%%%%%%%%%%%%%%%%%%%%%%%%%%%%%%%
\subsection{Advection-Diffusion Dynamics with RANS Equations}
In order to obtain the concentration of the emitted molecules in the environment, we need to solve the advection-diffusion equation.
\begin{equation}\label{e1_adv_diff}
   \frac{\partial {c}}{\partial t}=\nabla \cdot ({D_{\varepsilon}} \nabla c)-\nabla \cdot (\vec{v}c),
\end{equation} where $c$ is the concentration and ${D_{\varepsilon}}$ is the eddy diffusivity coefficient of the water molecules. $c_{0}$ is the amount of the molecules which are released into the channel at $t=0$, and $v$ is the velocity field of the environment flow. Generally, there are two restrictions in solving {\eqref{e1_adv_diff}}. First of all, $\vec{v}$ is a function of the space and time which means that in any arbitrarily location and time, the velocity components should be calculated and substituted in {\eqref{e1_adv_diff}} in order to find concentration distribution. In the literature \cite{unluturk2017end}, this restriction has been ignored and they considered the velocity field constant spatially to find a closed-form relation for the concentration distribution. Secondly, the eddy diffusivity, ${D_{\varepsilon}}$, will be changed as the messenger molecules (MMs) go far away from the transmitter and it is not isotropic. In the literature \cite{farsad2016comprehensive}, the eddy diffusivity mostly has been considered isotropic which means that the information particles in the channel can be dispersed in any directions equivalently whilst this assumption is not accurate due to the essence of the turbulent flow \cite{roberts2002turbulent}.  

Based on the discussed restrictions, considering anisotropic velocity and eddy diffusivity and also, considering time-variant velocity simultaneously makes the problem complicated and finding a closed-form solution is almost impossible. In order to address the foregoing problem, the velocity distribution should be obtained and employed in {\eqref{e1_adv_diff}}. One of the scheme to obtain the velocity distribution is using the numerical packages to simulate the flow field and solve the \textbf{Reynolds-Average-Navier-Stokes (RANS)} equations \cite{roberts2002turbulent}. The key characteristic of the numerical packages like COMSOL Multiphysics is that they solved the RANS equations with the mass transport equation \eqref{e1_adv_diff} simultaneously and it considers the effects of eddies on transporting the molecules from transmitter to receiver.
%%%
\begin{equation}
\begin{split}
\label{RANS}
c\overline{u}_{j}\frac{\partial \overline{u}_{i}}{\partial x_{j}} = c\overline{f}_{i} 
\!+\! \frac{\partial}{\partial x_{j}} \bigg[  \!-\overline{p}\delta_{ij} \!+\! \mu\bigg(\! \frac{\partial \overline{u}_{i}}{\partial x_{j}} \!+\! \frac{\partial \overline{u}_{j}}{\partial x_{i}} \!\bigg) \!- c\overline{u_{i}'u_{j}'}\bigg],
\end{split}
\end{equation} where $c$ represents density or concentration which depends on a number of pressure, velocity, and sheer stress gradients. The dynamic viscosity of the fluid is $\mu$, and the term $c\overline{u}_{j}\frac{\partial \overline{u}_{i}}{\partial x_{j}}$ represents the change in mean momentum of the fluid element due to the unsteadiness in the mean flow and the convection by the mean flow. This is balanced by the mean body force $\overline{f}_{i}$, the isotropic stress from the pressure field $\overline{p}\delta_{ij}$, the viscous stresses, and the apparent stress $-c\overline{u_{i}'u_{j}'}$ owing to the fluctuating velocity field (Reynolds stress). Whilst there are statistical approximate solutions in the form of eddy diffusivity, general tractability is still a challenge for modeling turbulent diffusion and that is why finite-element simulation is used.\\

%%%%%%%%%%%%%%%%%%%%%%%%%%%%%%%%%%%%%%%%%%%%%%%%%%%%
\subsection{Non-intrusive PCE}
We employed the non-intrusive PCE since it does not need any changes in the built-in code of COMSOL Multiphysics solver, to add uncertainty in a set of parameters including three components of inlet velocity ($\textbf{V}=u_0i+v_0j+w_0k$) and initial concentration ($c_0$). These parameters without uncertainty should be ($\textbf{V}=4i+0j+0k$) and $\SI{4}{\mol/\meter^3}$, repectively. 
%%%%%%%%%%%%%%%%%%%%%%%%%%%%%%%%%%%%%%%%%
\begin{figure*}[t]
	\centering
	\includegraphics[width=0.9\linewidth]{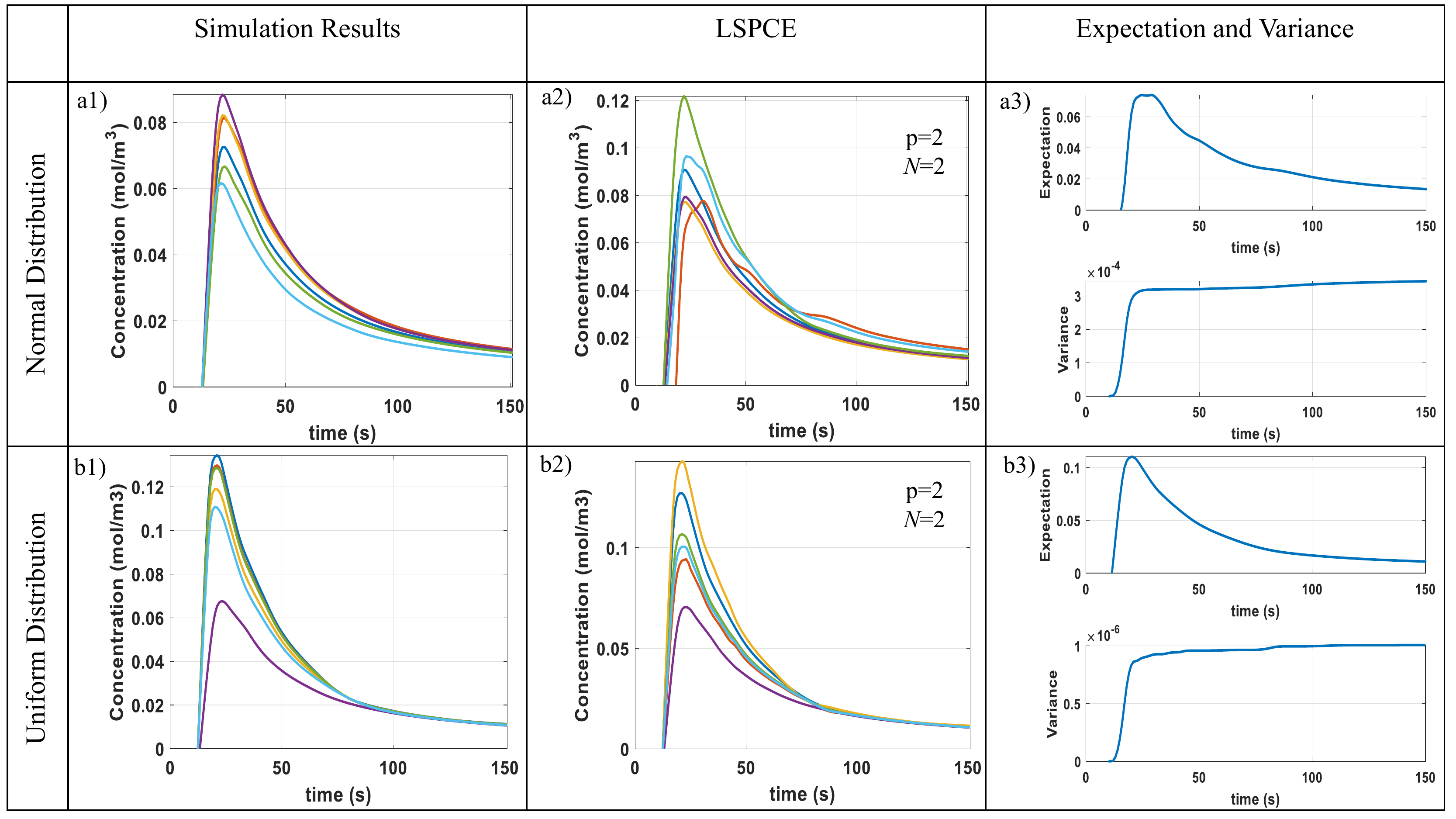}
	\caption{Simulation results and corresponded LSPCE with expectation and variance for normal distribution and uniform distribution.}
	\label{4}
\end{figure*}
%%%%%%%%%%%%%%%%%%%%%%%%%%%%%%%%%%%%%%
\begin{figure*}[t]
	\centering
	\includegraphics[width=0.9\linewidth]{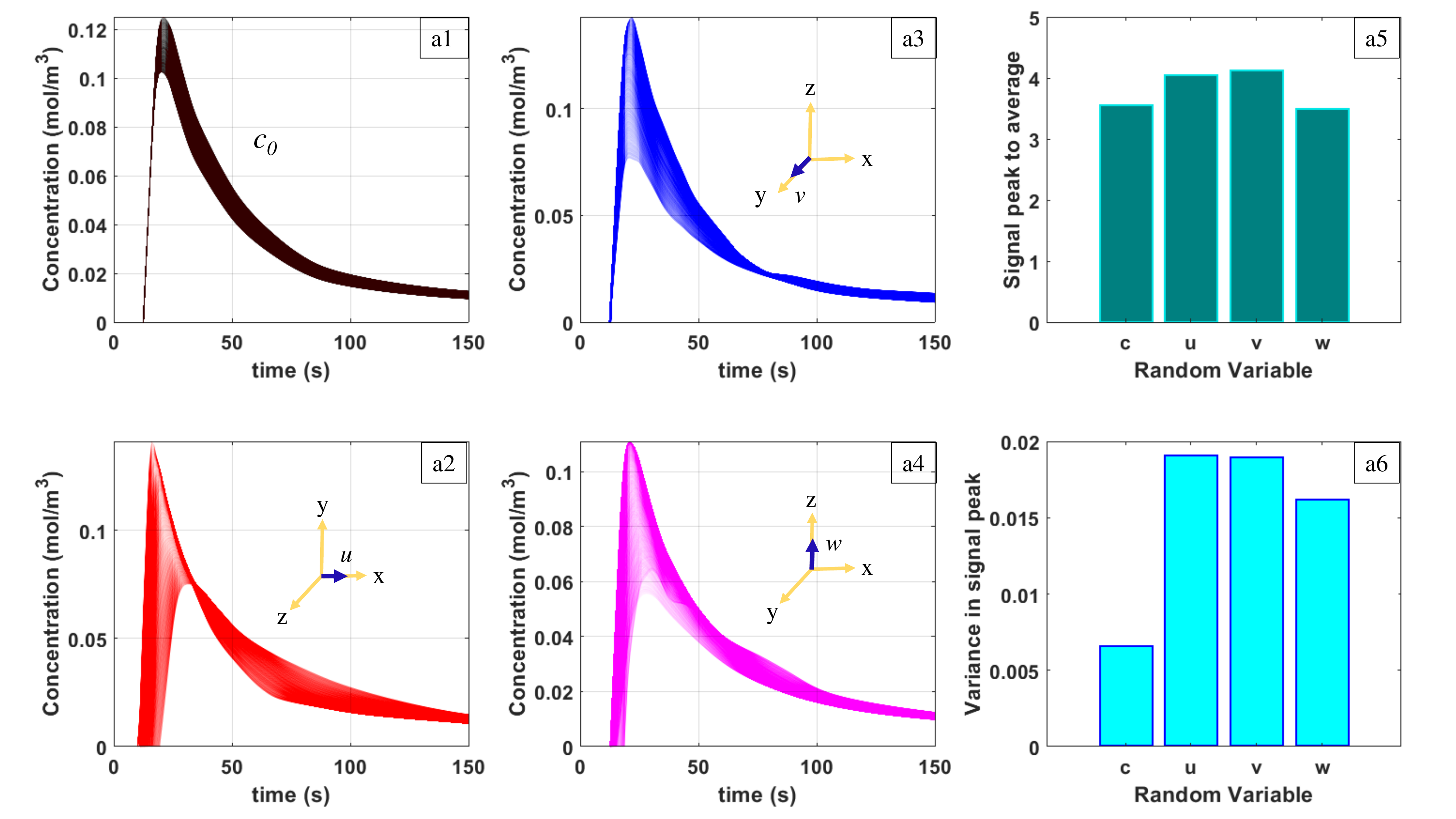}
	\caption{10000 concentration profiles obtained by LSPCE for input random variable of (a1) $c_0$ (a2) $u_0$, (a3) $v_0$ (a4) $w_0$. (a5) Signal peak to average and (a6) Variance in signal peak  for all of 10000 concentration profiles for each input random variable.}
	\label{5}
\end{figure*}
%%%%%%%%%%%%%%%%%%%%%%%%%%%%%%%%%%%%%%%%
For building the PCE, we need to truncate the expansion. Generally, there are two ways of truncating the PCE: 1) truncating the series up to an order of interest. For example, if the order of interest is $k$, then for the series of \eqref{e11}, we have:
\begin{equation}\begin{split}
\label{e17}
  \phi(\Xi_1,\Xi_2) = \sum_{0\leq(n+m)\leq k} a_{m,n}\, P_{1,m}(\Xi_1) P_{2,n}(\Xi_2).
\end{split}\end{equation}
2) We can consider all possible combinations of the order for each univariate polynomial. This method which is called \textbf{tensor product truncation} can be shown as:

\begin{equation}\begin{split}
\label{e18}
  \phi(\Xi_1,\Xi_2) = \sum_{0\leq n\leq k}  \sum_{0\leq m\leq k} a_{m,n}\, P_{1,m}(\Xi_1) P_{2,n}(\Xi_2).
\end{split}\end{equation}

In this paper, we utilized the first method since it only considers the terms in the series which has higher order than the other terms \cite{owen2017comparison}. If we consider the number of the random variables $n$ and the maximum order of the applied polynomial $p$, then the number of the terms in the truncated PCE (or size of experimental design) would be $N$=$n+p \choose p$ for the first method of truncating, and for the tensor product truncation that would be $N=(p+1)^n$ \cite{owen2017comparison}. After generating $N$ random numbers (we should use Legendre polynomials if we choose the random numbers from uniform distribution and use Hermite polynomial if the normal distribution has been used), the deterministic COMSOL code should be evaluated at these random numbers to obtain the left hand side of the \eqref{e14}. Then, the polynomials in the right hand side of \eqref{e14} are also evaluated at these points and finally, by solving a linear regression problem, the $a_n$ coefficients will be obtained. Now, we can construct the expansion and use it as an approximation formula without running the solver for a new set of random numbers. What is crucial in regression is the size of the experimental design or $N$. If we evaluate the PCE exactly at $N$ random numbers, it does not yield to a stabilized series and we will see oscillation in the established PCE (see Fig.\ref{2}-a2, -b2, -c2). To overcome such a problem, we can choose the size of the experimental design 2$\times N$ as suggested by many authors in literature \cite{hosder2007efficient, owen2017comparison}. So, since we evaluate the series in 2$\times N$ random numbers, we should utilize the least squares  to solve the regression problem (this method is called \textbf{Least Square Polynomial Chaos Expansion (LSPCE)}). By doing so, the oscillation will be removed and we will have a stabilized series (see Fig.\ref{2}-a3, -b3, -c3).

%%%%%%%%%%%%%%%%%%%%%%%%%%%%%%%%%%%%%%
\section{Results \& Discussion} 
%%%%%%%%%%%%%%%%%%%%%%%%%%%%%%%%%%%%%
\subsection{Convergence test of PCE}
Generally, since the input random variables are fixed based on each problem of interest, the order of the polynomial has to be changed to reach to the convergent coefficients. If we see Fig. \ref{2}-a3, -b3, -c3, we cannot differentiate which order of $p$ yields to convergence of PCE and it is hard to assess without considering the statistic of the random outputs. So, we should calculate the expectation and variance of the concentration in these plots from \eqref{e7} and \eqref{e8}, respectively. 
We can see that the expectation for $p=2$ and $p=3$ has been overlapped whilst for $p=1$ with the other cases this has not happened (see Fig. \ref{3}-a1). For illustrating this issue more clearly, we compared the absolute difference of the expectation between $p=2-3$ and $p=2-1$ in Fig. \ref{3}-a2. The order of difference between these two cases is 0.1 which demonstrates that $p=1$ cannot be the appropriate polynomial order. Then, we tracked the variance by increasing the polynomial order and we can see that by increasing from $p=2$ to $p=3$, the variance changes only one order, but by increasing from $p=1$ to $p=2$, the variance changes two order. Thus, $p=2$ is the appropriate polynomial order and choosing order more than two only builds up the simulation time with minor effects on improving the accuracy of the results.
\subsection{Comparing different PDFs in PCE}
Based on the type of the inputs, different PDFs should be employed to get the most reliable results. For clarifying this issue, we considered the dynamic viscosity of the water ($\mu$) and the turbulent Schmidt number ($Sc$) as the input random variables (dynamic viscosity is uncertain since it changes based on the environment temperature and it is $8.9\times 10^{-4}$ Pa.s in $25^\circ$C, and the turbulent Schmidt number is uncertain since it is determined experimentally and it is usually considered 0.71) with normal and uniform distributions (see Fig. \ref{4}) . The concentration profiles of uniform PDF are more straight than normal ones (see Fig.\ref{4}-a2 and -b2) as far as the normal distribution is unbounded whilst the uniform distribution is bounded (between -1 and 1). So, for input parameters like dynamic viscosity of the water that are order of $10^{-4}$, considering unbounded PDF would not yield to a reliable results. This subject is also vivid in the expectation and variance (see Fig.\ref{4}-a3 and -b3). The variance for uniform PDF is two orders less than the normal PDF which demonstrates that the input random numbers are mostly around $8.9\times 10^{-4}$ Pa.s.
%%%%%%%%%%%%%%%%%%%%%%%%%%%%%%%%%%%%%%%%%%%%%

\subsection{Statistical results}
\subsubsection{Multiple uncertainties}
In Fig. \ref{5}, we employed the LSPCE to investigate the statistical properties of having one uncertain parameter in TDMC channel. In Fig.\ref{5}-a1 the uncertain parameter is $c_0$ and in Fig. \ref{5}-a2 to -a4 the uncertain parameters are different components of the inlet velocity ($\textbf{V}=u_0i+v_0j+w_0k$). We run six simulations per parameter ($\frac{(n+p)!}{n!p!}=\frac{(1+2)!}{2!}=3\times2=6$ LSPCE) to build the series and then we exploited that series for calculating the concentration of 10000 uniform random inputs. If we wanted to do simulation to get these 10000 profiles, it would be very time-consuming whilst by using PCE, it took less than ten minutes to get the concentration profiles. 

\subsubsection{Communication performance impact}
Fig.\ref{5}-a1-a4 also illustrates the density of concentration profiles for each input random variable. We can see that the uncertainty in the input mostly reflected in the peak of the channel responses and the tail of the channel response is immune from the input uncertainty. Therefore, we can conclude that only the signal strength is affected and not so much the inter-symbol-interference (ISI).

Also, we can see that the channel response is more sensitive to the alteration of the inlet velocity compared to initial concentration. This is very crucial point in a sense that in designing the communication transmitter experimentally, more attention should be devoted to cylinder and piston shape compared to the amount of initial molecules.

The received signal peak-to-average ratio is also important from a power amplifier linearity perspective, which needs to convert molecular count into a received signal. Fig.\ref{5}-a5 illustrates that the ratio does not face a significant change by introducing uncertainty in the initial concentration or inlet velocity, and Fig.\ref{5}-a6 which displays the variance in signal peaks, demonstrates that uncertainty in the initial concentration leads to more noise in the channel compared to uncertainty in the inlet velocity.

\section{Conclusion \& Future Work}

Currently we do not understand how uncertainty or noise in a variety of parameters affect the received signal concentration, and nor do we have an analytical framework to tackle this challenge. In this paper, we utilize Polynomial Chaos Expansion (PCE) to show to uncertainty in parameters propagates to uncertainty in the received signal. The PCE method has a significant time saving compared to Monte-Carlo simulations and offers theoretical insight. The analytical results are validated using multi-physics COMSOL simulations in a Turbulent Diffusion Molecular Communication (TDMC) channel. Our uncertain parameters are initial concentration, injection velocity, dynamic viscosity of the water, and turbulent Schmidt number. We demonstrated that how the uncertainty in the aforementioned parameters propagates through the channel and can affect the received signal response. 

The research conducted here in PCE and uncertainty propagation can pave the way for future information theoretic insights, as well as guide experimental design. In the future, we will focus on understanding the channel capacity as a function of uncertainty. \\

\IEEEpeerreviewmaketitle

%%%%%%%%%%%%%%%%%%%%%%%%%%%%%%%%%%%%%%%%%%%%%%%%%%%%
%%%%%%%%%%%%%%%%%%%%%%%%%%%%%%%%%%%%%%%%%%%%%%%%%%%%
%%%%%%%%%%%%%%%%%%%%%%%%%%%%%%%%%%%%%%%%%%%%%%%%%%%%
%%%%%%%%%%%%%%%%%%%%%%%%%%%%%%%%%%%%%%%%%%%%%%%%%%%%
%%

\bibliographystyle{IEEEtran}
\bibliography{turbulent_diffusion}

\end{document}